%% file: creative_destruction.tex
\documentclass[12pt]{article}

\usepackage[letterpaper,margin=1in]{geometry}
\usepackage{times}
\usepackage[T1]{fontenc}
\usepackage[utf8]{inputenc}
\usepackage{setspace}
\usepackage{graphicx}
\usepackage{booktabs}
\usepackage{dcolumn}
\usepackage{multirow}
\usepackage{array}
\usepackage{threeparttable}
\usepackage{float}
\usepackage{ragged2e}
\usepackage{amsmath,amssymb}
\usepackage[numbers,round,sort&compress]{natbib}
\usepackage{caption}
\usepackage{tikz}
\usetikzlibrary{arrows.meta, positioning, calc, decorations.pathreplacing}
\usepackage[hidelinks]{hyperref}
 
\providecommand{\sym}[1]{\ifmmode^{#1}\else\(^{#1}\)\fi}

\newcommand{\subhead}[1]{\subsection*{\normalsize\mdseries\itshape #1}}

\newif\ifSM \SMtrue

\input{figures/numbers_paper.tex}

\begin{document}

\begin{flushleft}
{\Large\bfseries Creative destruction: Human-AI competition in an experimental market}

\bigskip

John J. Conlon$^{1}$ and Peter Schwardmann$^{2}$

\medskip

\noindent $^{1}$Department of Economics, The Ohio State University, Columbus, OH 43210, USA.\\
$^{2}$Department of Social and Decision Sciences, Carnegie Mellon University, Pittsburgh, PA 15213, USA.

\bigskip

\end{flushleft}

\bigskip

\noindent\textbf{Abstract.} We study two hypotheses for how a human role in production could persist despite growing AI capabilities. Humans might have a durable advantage in taste, designing better products even as AI agents automate routine tasks, or consumers might intrinsically prefer that humans have a hand in production. In our experiment, participants act as producers, designing computer games with an AI coding agent, and as consumers who buy them. Across three studies (N = 1{,}025), we find no evidence for either hypothesis. Games designed by an AI on its own outperform those of human producers, including both amateurs and professional creatives. Human producers do not anticipate the AI's advantage. Revealing whether a game is human-designed does not raise consumers' willingness to pay for the product.
\newpage

\subsection*{Introduction}

Economic production increasingly involves collaboration between humans and artificial intelligence (AI) systems. The share of US workers reporting using AI on the job doubled to 40\% between 2023 and 2025, and its uptake in the workforce has been faster than that of the personal computer or the internet \citep{appel2025economicindex, bick2024rapid}. Firms deploy large language models (LLMs) and other AI systems for tasks as varied as coding, job interviewing, customer support, clinical documentation, medical imaging, management consulting, sports umpiring, and debt collection \citep{appel2025economicindex, brynjolfsson2025generative, jabarian2026voice, lukac2025ambient, agarwal2023combining, dellacqua2023navigating, peng2023impact, almogetal2024oversight, choi2025ai}, and human employees spend an increasing share of their time managing and directing AI agents \citep{humlum2025still, iscenko2026google, codreanu2026ai, hitzig2026agentic}. These expanding capabilities have enormous potential upsides \citep{brynjolfsson2025generative, noy2023experimental, peng2023impact, dellacqua2023navigating}, but they also raise concerns about AI eventually displacing human labor. In contrast to earlier technological disruptions, LLMs increasingly perform tasks that once seemed quintessentially human: communication, art, reasoning, and knowledge work more broadly. AI automation therefore poses not only a financial risk to workers whose jobs it might displace \citep{eloundou2024gpts, hui2024short, kim2026does}, but a threat to a primary source of meaning and pride in people's lives.

Two hypotheses are commonly advanced for why a human role in production might persist even as AI capabilities grow. First, on the supply side, humans might have a durable advantage in \textit{taste}: they may be better at knowing, or more creative in designing, the products that other humans will prefer \citep{chiang2024art, chakrabarty2024art}. If so, while AI agents will increasingly perform more routine tasks (e.g., coding), human guidance will remain crucial in directing them toward productive ends \citep{autor2015why, agrawal2019exploring}. Some commentators go even further, arguing that by removing technical barriers through automation, AI can unlock the creative and productive potential of a much larger set of workers  \citep{reuters2023huang, autor2024applying, roose2025coder}. Second, on the demand side, consumers might intrinsically prefer that humans retain a hand in production \citep{fuchs2015handmade, granulo2021symbolic, bellaiche2023humans, morewedge2022preference, imas2026scarce}. That is, human labor might remain integral to the production process either for instrumental reasons---because they contribute to the creation of better products---or for intrinsic reasons, because consumers demand human-made products \textit{per se}.

Testing these hypotheses experimentally requires constructing a market where human producers---collaborating with AI agents---compete with entirely AI systems, and where we can manipulate what consumers know about the human involvement (or lack thereof) in the production process. We construct such a market in three preregistered studies that feature the production of, and decisions over whether to purchase, more than \nUniqueValuedGamesFloor{} unique computer games (which can be explored at \href{https://human-ai-games.github.io/}{https://human-ai-games.github.io}). Our experimental market then allows us to answer three research questions:

\begin{itemize}\setlength{\itemsep}{0pt}
\item[] \textbf{RQ1:} Do human producers, whether amateurs or professionals, have an advantage over AI in taste?
\item[] \textbf{RQ2:} Do producers believe that they have such an advantage?
\item[] \textbf{RQ3:} Do consumers intrinsically prefer human-made products?
\end{itemize}

\noindent RQ1 and RQ3 test our two main hypotheses. RQ2 asks whether producers themselves believe the first hypothesis, and therefore whether its failure would come as a surprise to them. Crucially, all key choices in our market are incentivized: producers earn what consumers are willing to pay for their games, consumers pay for the games they choose to play, and producers' predictions are rewarded for accuracy.

\subsection*{Experimental design}

\begin{figure}[t!]
\centering
\definecolor{mpblue}{RGB}{31,119,180}
\definecolor{mporange}{RGB}{255,127,14}
\definecolor{mpgreen}{RGB}{44,160,44}
\begin{tikzpicture}[
  font=\small,
  every node/.style={inner sep=3pt},
  phase/.style={
    rectangle, rounded corners=2pt, draw=black!70, thick,
    minimum height=0.9cm, minimum width=1.15cm,
    align=center, font=\footnotesize\linespread{0.9}\selectfont,
  },
  ptask/.style={phase, fill=mpblue!20,   draw=mpblue!80!black},
  ctask/.style={phase, fill=mpgreen!20,  draw=mpgreen!70!black},
  aux/.style={phase, fill=black!5},
  aion/.style={
    rectangle, rounded corners=2pt, draw=mporange!80!black, thick, dashed,
    minimum height=0.75cm, minimum width=1.5cm, align=center,
    font=\footnotesize\linespread{0.9}\selectfont, fill=mporange!20,
  },
  flow/.style={-{Latex[length=1.6mm,width=1.4mm]}, thick, black!60},
  hook/.style={-{Latex[length=1.6mm,width=1.4mm]}, thick, dashed, mporange!80!black},
]
\node[aux] (intro) {Instructions};
\node[aux, right=0.55cm of intro] (demo) {Demo Phase};
\node[ptask, right=0.55cm of demo] (p1) {P1};
\node[ptask, right=1.55cm of p1]   (p2) {P2};
\node[ptask, right=1.55cm of p2]   (p3) {P3};

\foreach \a/\b in {intro/demo, demo/p1, p1/p2, p2/p3}
  \draw[flow] (\a) -- (\b);

\coordinate (gap12) at ($(p1.east)!0.5!(p2.west)$);
\coordinate (gap23) at ($(p2.east)!0.5!(p3.west)$);

\node[aion, above=1.2cm of p2] (ai) {AI-Designed game};

\draw[hook] (ai.south) --
     node[midway, above left=-2pt, font=\scriptsize, text=mporange!70!black] {(33\%)}
     (gap12);
\draw[hook] (ai.south) --
     node[midway, above right=-2pt, font=\scriptsize, text=mporange!70!black] {(33\%)}
     (gap23);

\draw[decorate, decoration={brace, amplitude=4pt, mirror},
      black!55, thick]
      ($(p1.south west)+(0,-0.15)$) -- ($(p3.south east)+(0,-0.15)$)
  node[midway, below=4pt, font=\small\bfseries, text=black!75]
    {Producer Phase};

\node[ctask, below=1.8cm of intro.south west, anchor=north west] (c1) {C1};
\node[ctask, right=0.35cm of c1]   (c2) {C2};
\node[ctask, right=0.35cm of c2]   (c3) {C3};
\node[ctask, right=0.35cm of c3]   (c4) {C4};
\node[aux,   right=0.55cm of c4]   (end) {Post-Experiment\\Survey};

\foreach \a/\b in {c1/c2, c2/c3, c3/c4, c4/end}
  \draw[flow] (\a) -- (\b);

\draw[decorate, decoration={brace, amplitude=4pt, mirror},
      black!55, thick]
      ($(c1.south west)+(0,-0.15)$) -- ($(c4.south east)+(0,-0.15)$)
  node[midway, below=4pt, font=\small\bfseries, text=black!75]
    {Consumer Phase};

\draw[flow, rounded corners=8pt]
  (p3.east) -- ++(0.5,0) |- ($(c1.north)+(0,0.5)$) -- (c1.north);

\coordinate (topP) at ($(ai.north)+(0,0.1)$);
\coordinate (botP) at ($(p3.south)+(0,-0.75)$);
\coordinate (topC) at ($(c1.north)+(0,0.6)$);
\coordinate (botC) at ($(c1.south)+(0,-0.75)$);
\coordinate (xA) at ($(p3.east)+(1.0,0)$);
\coordinate (xB) at ($(p3.east)+(2.6,0)$);
\draw[decorate, decoration={brace, amplitude=5pt}, black!55, thick]
  (xA |- topP) -- (xA |- botP)
  node[midway, right=6pt, align=left, font=\small\bfseries, text=black!75] {Study 2};
\draw[decorate, decoration={brace, amplitude=5pt}, black!55, thick]
  (xA |- topC) -- (xA |- botC)
  node[midway, right=6pt, align=left, font=\small\bfseries, text=black!75] {Study 3};
\draw[decorate, decoration={brace, amplitude=6pt}, black!55, thick]
  (xB |- topP) -- (xB |- botC)
  node[midway, right=7pt, align=left, font=\small\bfseries, text=black!75] {Study 1};

\end{tikzpicture}
\caption{\textbf{Experimental design.}
In Study 1 producers design three computer games (P1--P3) before stating their willingness to pay for up to four games (C1--C4) in the consumer phase. A randomly selected two-thirds of participants watch an AI design a game with no human input, after either their first or their second design task. In Study 2, professional producers engage only in the production phase. In Study 3, participants complete only a consumer phase with games produced in Studies 1 and 2.}
\label{fig:1}
\end{figure}
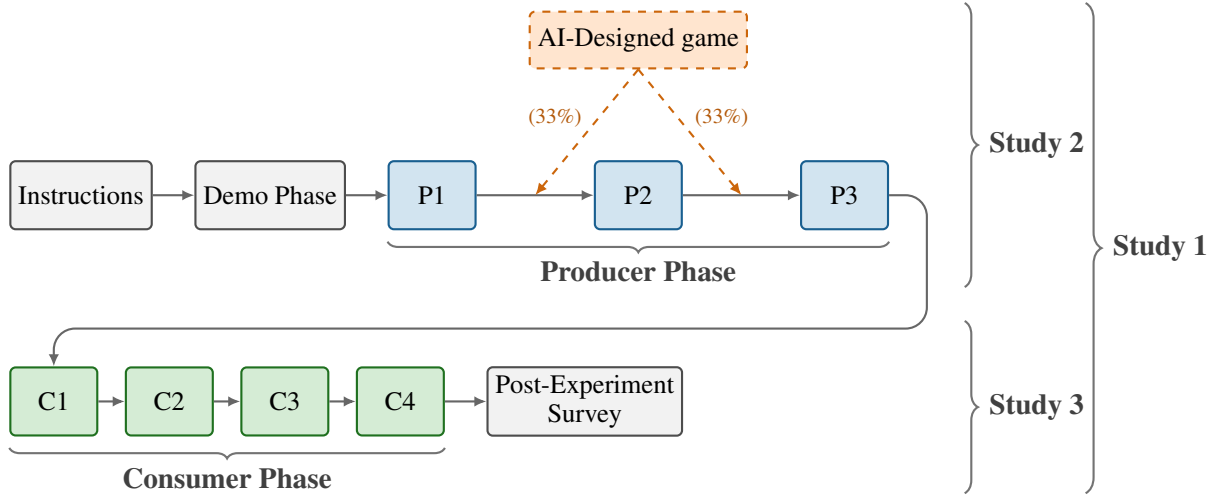

Our three studies involve 1,025 US-based participants recruited from Prolific, an online research platform (296 in Study 1, 180 in Study 2, and 549 in Study 3; summary statistics in table~\ref{tab:S1}). As shown in Fig.~\ref{fig:1}, Study 1 contains the full market with participants acting first as producers and then as consumers of computer games. Participants in Study 2 were professionals in creative industries who engaged only in the production phase. Participants in Study 3 only made consumption decisions. See supplementary materials for full methods.

\begin{figure}[t!]
    \centering
    \includegraphics[width=\textwidth]{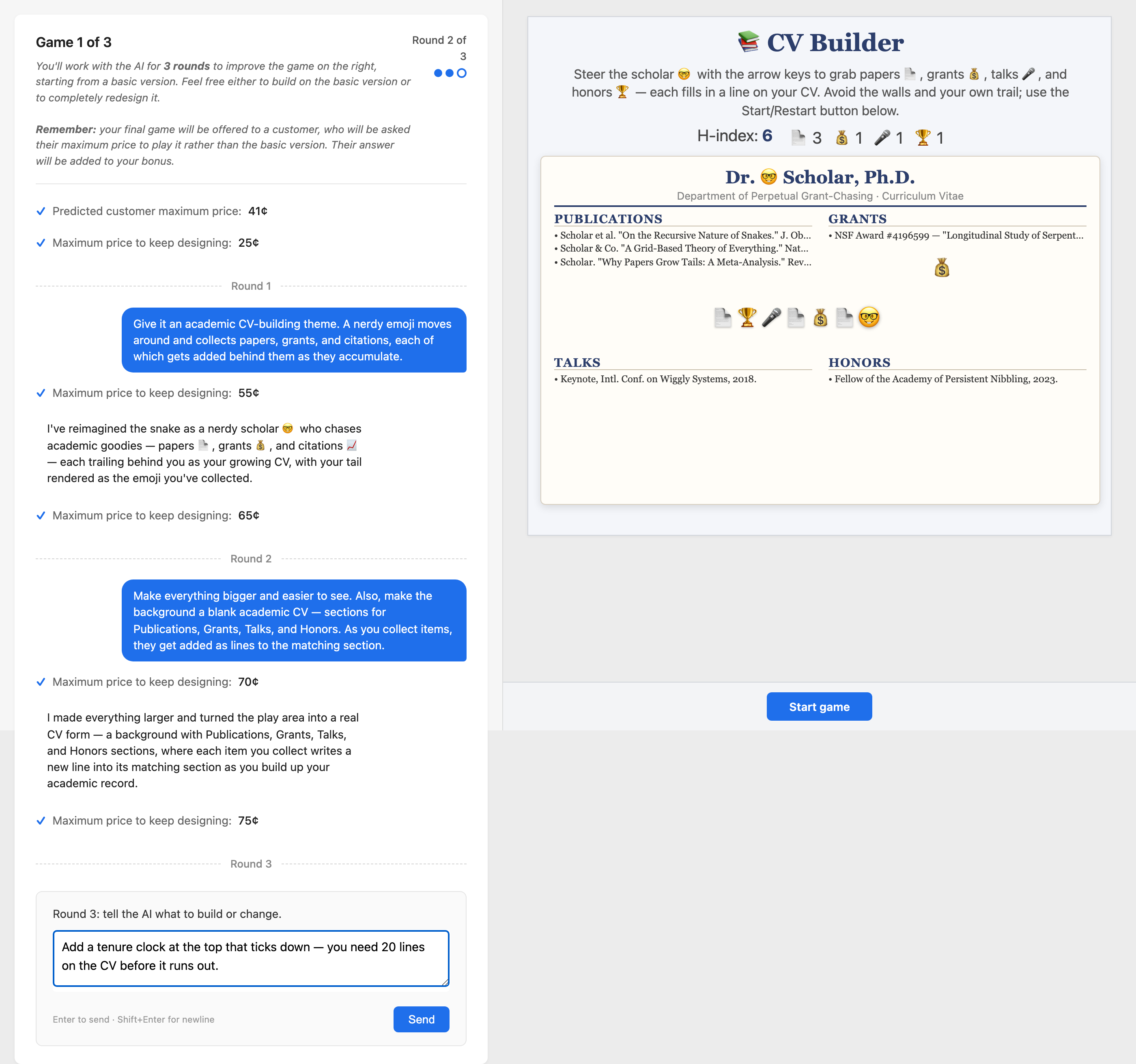}
    \caption{\textbf{The production interface.} A design task in the producer phase. The current version of the game appears on the right; messages to the coding agent and the elicitations asked of the participant appear on the left.}
    \label{fig:2}
\end{figure}

\subhead{Study 1: Participants as producers and consumers}
As producers, participants first designed three games in collaboration with an AI coding agent powered by a large language model (Anthropic's Opus 4.8), accessed live via an interface embedded into the experiment. Each task started with a plain black-and-white ``basic game''---for example, falling blocks resembling Tetris, or a paddle game similar to Pong---which participants improved through three rounds of messages to the agent. Upon receiving the participant's message, the coding agent would modify the code for the game, the updated version of which appeared on the screen, replacing the previous version. Between rounds, participants tested out the current state of the game to develop ideas for how to change it. At various points during the task, a participant's willingness to continue designing the game and their prediction of the price it will fetch on the market were elicited. Fig.~\ref{fig:2} provides an example of the experimental interface in the production phase.

After the production phase, participants entered the consumer phase of the study. After testing another participant's game and the basic game it came from, participants were asked their willingness to pay (WTP), using an incentive-compatible payment rule \citep{becker1964measuring}, to play the other participant's game rather than the basic game for at least one minute. They made consumption decisions for up to four games, and their stated WTPs constituted the market incentives producers faced: a consumer's WTP for a game was paid to its producer as a bonus.

Two-thirds of producers were randomized to see an AI design a game on its own, without human input, either after their first or second design task. These producers were truthfully told that we prompted an LLM (Anthropic's Sonnet 4.6) to redesign the same basic game they had just seen ``to be much more creative, engaging, and fun to play,'' and that its single response was passed to the same coding agent they had just interacted with. Participants then saw the LLM ``type'' its message into a text box to be sent to the coding agent, before getting to test the game the coding agent created in response (see Fig.~\ref{fig:S1} for an example). The AI-designed games then entered the pool of games the consumer phase drew from alongside the human-designed games. 

Comparing consumers' WTP for human- and AI-designed games addresses RQ1. To address RQ2, we asked producers to predict consumers' WTP for their own games and for the AI's, both before and after they had witnessed the AI design a game.

Finally, we randomized whether consumers were told how games were produced. In the Blind arm, they were given no information, while in the Disclosed arm they were told whether the game was designed ``by another participant who worked with an AI, just like you did'' or ``by an AI on its own''. Comparing the difference between WTPs for AI- and human-made games across the two disclosure arms addresses RQ3. In particular, an intrinsic preference for human-made products would manifest as a greater advantage or smaller disadvantage of human-made games in the Disclosed arm of the experiment.

\subhead{Study 2: Professional producers}
 In Study 2, we recruit 180 participants with jobs in creative industries whose experience might boost their performance. Of these, \nProGames{} are game developers or work in the video game industry, \nProGraphic{} are graphic designers, and \nProContent{} are content creators, identified through Prolific's occupational screeners. In a post-experiment survey, \creatorJobPct\% of them report that their job has recently involved them ``creating something for other people to use, look at, read, or play — for example writing, visual design, a game, or software''. We therefore refer to these participants as ``professionals.'' The professionals completed only the production phase of the experiment.

\subhead{Study 3: Dedicated consumers}
In Study 3, a separate 549 participants completed only the consumer phase of the experiment. The pool of games these consumers faced comprised the 503 games created by the professionals in Study 2, 388 ``amateur'' games made by Study 1's participants, and 297 AI-designed games from both prior studies. Study 3 participants also rated each game on the dimensions of fun, creativity, polish, and clarity on 1--9 Likert scales. Table~\ref{tab:S2} shows that these ratings are highly predictive of participants' WTP.

The consumers in Study 3 were randomized into three treatments. First, a Blind treatment did not disclose whether each game was designed by a human or AI, though all participants were aware that they might be shown games of both types. Second, a Disclosed + Paid treatment told participants whether each game was AI- or human-designed, and for the latter made salient that their WTP would be paid to the producer. Finally, a Disclosed + Unpaid treatment also told participants whether a game was AI- or human-designed, but participants were (truthfully) told that their WTP would not affect the producer's payment.

\subsection*{Results}

\subhead{Human producers do a good job, and have a good job}
Before turning to our main research questions, we show that the job we created in our experiment is valued by producers and that the products they make are valued by consumers. It is, therefore, the type of productive human endeavor we may worry about losing as a society. The average consumer was willing to pay \$\WTPhuman{} to play a human-designed game rather than the basic game it came from, and \shareWTPpos\% of stated WTPs were strictly positive. These are meaningful sums: consumers only played games for roughly a minute on average, and the alternative to buying a human-designed game was itself playing a working game. As a point of comparison, the average US movie ticket costs $\$$16.30 in 2026 \citep{cabletv2026movie}, or about 18 cents per minute of a 90-minute film; consumers' willingness to pay per minute for our participants' games is about twice that. In this sense, participants did a good job of designing products that consumers valued.

Participants' job as a producer also delivered non-monetary rewards to them. We find that \prideMod\% of producers report being at least moderately proud of their game (\prideExt\% ``extremely'' so), and \enjoyMod\% at least moderately enjoyed designing it (\enjoyLoved\% ``loved'' it). Two incentivized questions corroborate these measures: \linkPos\% were willing to pay for a personal link to their game to keep or share after the experiment, and \sitePos\% of producers were willing to pay to have it posted on a public website for other participants. We also include an incentivized measure of producer surplus: participants' WTP to undertake the design task rather than skip it and forgo the bonus, which was positive for \surplusPos\% of producers and averaged \$\surplusMean{} at the beginning of the design task (table~\ref{tab:S3}). This amounts to approximately half the hourly wage participants earned in the experiment. We elicited this WTP to keep designing several more times throughout the process; it remains high and if anything rises over time (Fig.~\ref{fig:S2}), showing that producers remain engaged throughout the tasks. These results suggest that, within the limits of a short controlled experiment, the participants not only did a good job, but had one. 

\begin{figure}[t!]
\centering
\includegraphics[width=\textwidth]{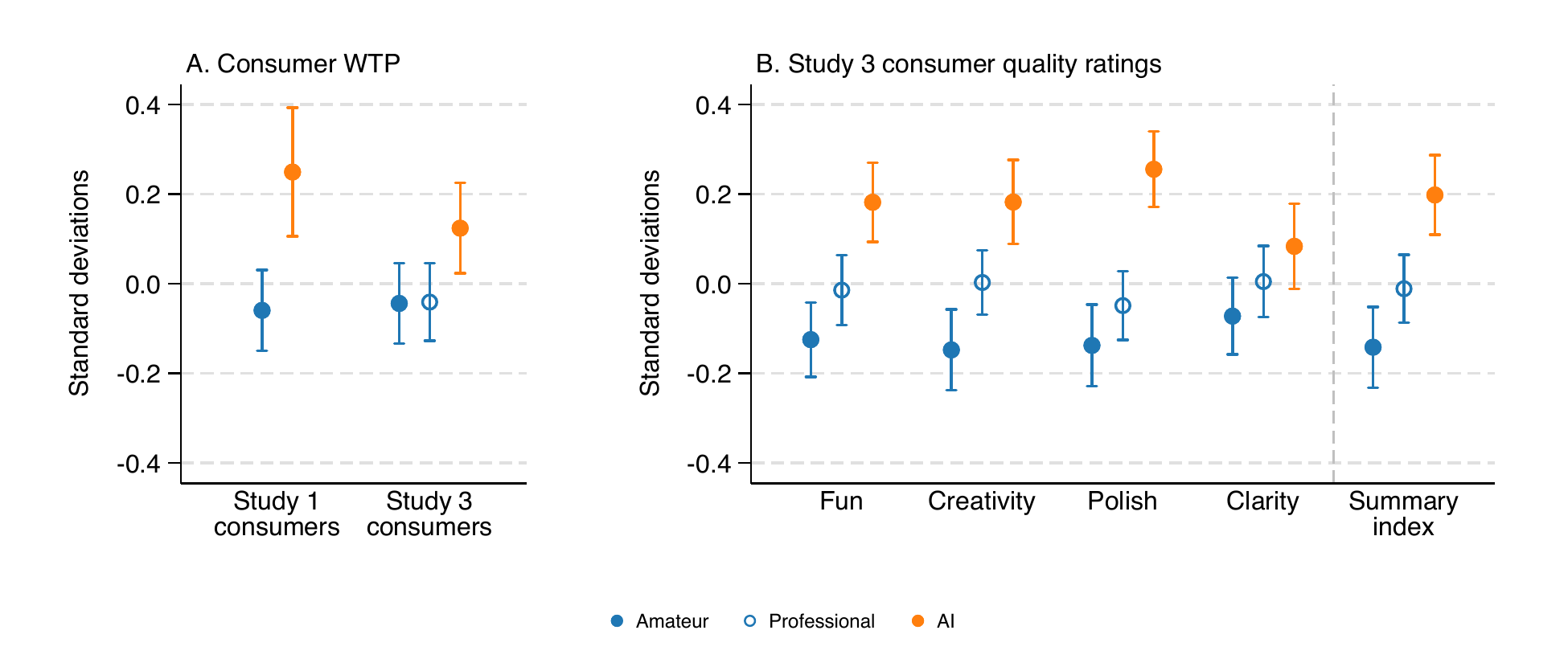}
\caption{\textbf{Consumer willingness to pay and quality ratings of amateur, professional, and AI-designed games.}
(\textbf{A}) Consumer WTP, separately for Studies 1 and 3. (\textbf{B}) Study 3 consumers' assessments of fun, creativity, polish, and clarity, and an inverse-covariance-weighted summary index of these four dimensions. WTP is standardized using the mean and standard deviation of all elicitations within each study; each quality dimension is standardized across all Study 3 elicitations. Dots show means, pooling disclosure treatments. Study 1 includes amateur and AI games; Study 3 includes amateur games from Study 1, professional games from Study 2, and AI games from both. Error bars are 95\% confidence intervals, with standard errors two-way clustered by consumer and producer.}
\label{fig:3}
\end{figure}

\subhead{AI-designed games outperform the games designed by both amateurs and professionals}
RQ1 asks whether human producers, whether amateurs or professionals, have an advantage over AI in taste and creativity. If so, we should find that human-designed games garner a higher WTP than those that the AI produces on its own. In fact, we find the opposite (Fig.~\ref{fig:3}A). Pooling across disclosure arms, consumers in Study 1 are willing to pay \$\WTPai{} on average for AI-designed games against \$\WTPhuman{} for human-designed games---a gap of \AIpremCents{} cents, or \AIpremSD{} standard deviations (SD) ($p\pAIprem$). Two features of the design are worth noting. First, producers and consumers were the very same people, and therefore producers could consult their own preferences to tailor the games they made to consumers (who by construction had the same preferences in expectation). Second, AI-only production was deliberately handicapped, as AI could not test out the game or iteratively improve it, and we only used a single prompt across all AI-designed games. Fig.~\ref{fig:3}A also shows that the AI's advantage carries over to the dedicated consumers of Study 3, and to the games of professional producers, whose games garner a \aiProWTPsd{} SD lower WTP than AI games on average ($p\pAiProWTP$).

Consumers' quality ratings suggest where the AI's advantage comes from (Fig.~\ref{fig:3}B). Study 3's consumers rate AI-designed games as more fun, more creative, and more polished than the games of amateurs and professionals alike. On our index combining the four ratings, professionals' games score \aiProIndex{} SD below AI games ($p\pAiProIndex$) and amateurs' games \aiAmIndex{} SD below ($p\pAiAmIndex$, significantly worse than professionals' games, $p\pProAmIndex$). Strikingly, consumers value AI games for their creativity, the dimension in which we might have expected humans to hold an advantage%
. One potential explanation is that coming up with ideas and judging whether they will improve a product is costly and difficult for humans, whereas an AI can ideate immediately and cheaply. The prompts sent to the coding agent are consistent with this channel: professionals' combined messages per game are longer than amateurs' (\wordsPro{} vs. \wordsAm{} words), and the AI's single messages are longer still (\wordsAI{} words). The AI may therefore outperform humans in part by generating more ideas, without compromising on their average quality.

Recall that two-thirds of producers had an opportunity to watch the AI design a game itself. A natural question is whether human producers can outperform the AI after they have had a chance to learn from it (e.g., by learning how best to prompt the coding agent). We do find some evidence of human games improving after watching the AI produce (we discuss this result further below), but even human games made after the producer saw the AI design a game garner a lower WTP than the AI games themselves, by \postExpoGapOne{} SDs in Study 1 ($p\pPostExpoGapOne$) and by \postExpoGapThree{} SDs in Study 3 ($p\pPostExpoGapThree$).

Finally, the AI's advantage is not confined to the middle or bottom of the WTP distribution. In Study 1, the AI--human gap in WTP is positive and its 95\% confidence interval excludes zero at every decile but the lowest; in Study 3, AI games outperform both amateur and professional games at every decile above the lowest (Fig.~\ref{fig:S3}), though effects are not all statistically significant. This result is noteworthy in the right tail of the distribution, where we might have expected hundreds of distinct human designers to generate more variance---and so better extremes---than one model sent the same prompt many times. Instead, we find no evidence for a standout group of human designers whose games beat the best of the AI's.

In answer to RQ1, we find no evidence that humans have an advantage over AI in taste. The AI's games outperform those of amateurs, of professionals, and of humans who have had the opportunity of learning from watching the AI design a game.

\begin{figure}[t!]
\centering
\includegraphics[width=\textwidth]{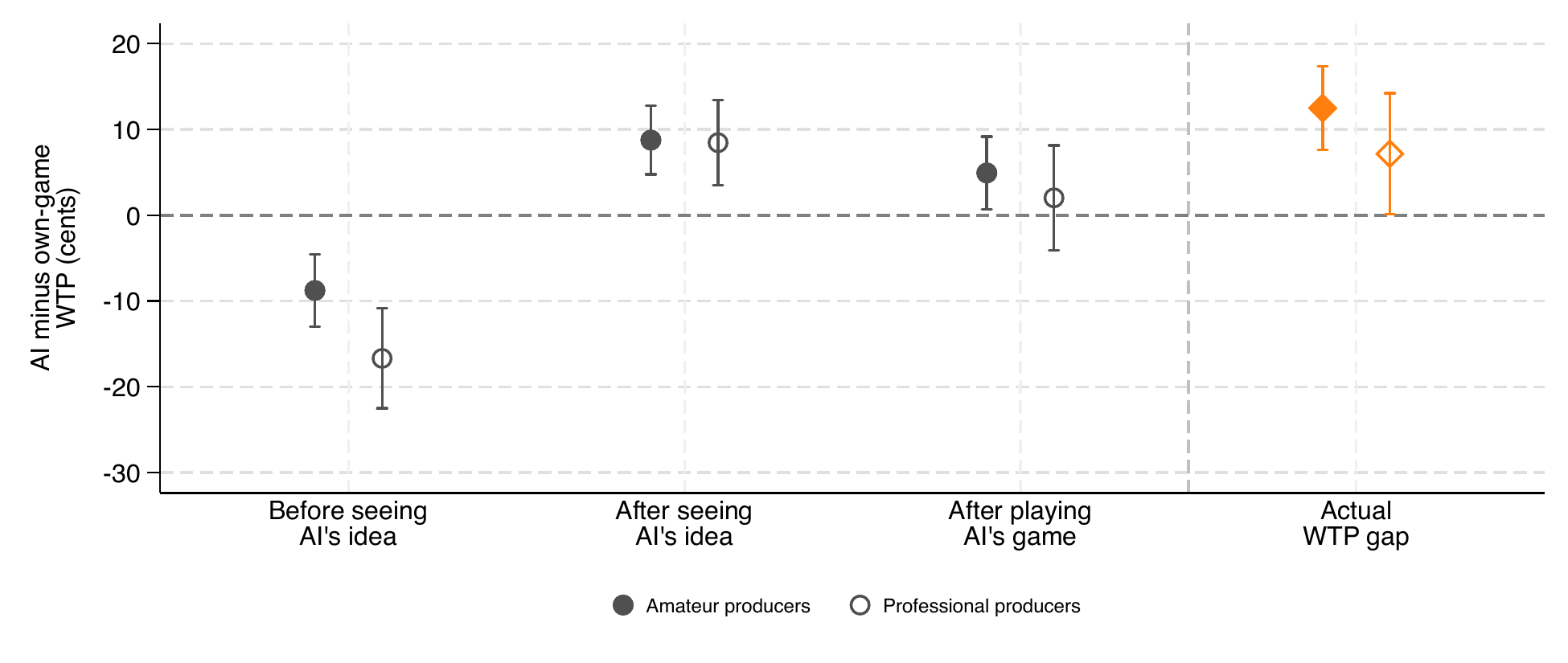}
\caption{\textbf{Predicted and actual willingness-to-pay gaps for matched games.}
Producers' predicted AI-minus-own-game WTP gaps before seeing the AI's idea, after seeing its idea, and after playing its game (gray circles), alongside the actual gap for the same games (orange diamonds). Filled symbols denote amateur producers (Study 1); hollow symbols denote professional producers (Study 2). The own-game prediction is producers' final prediction of the consumer's WTP for the game completed just before the AI-designed game. Actual WTP comes from Study 1 for amateurs, with the same consumer evaluating both games, and from Study 3 for professionals. Positive values indicate an AI advantage. Error bars are 95\% confidence intervals. Professionals' actual gap uses standard errors two-way clustered by consumer and producer; all other gaps use robust standard errors.}
\label{fig:4}
\end{figure}

\subhead{Producers expect to beat the AI}
RQ2 asks whether our participants believe that they have an advantage in taste. We ask producers after they have finished designing their game to predict consumers' WTP to play it. Producers also predict WTP for AI-designed games, both before and after they see the AI's design idea and its implementation. Fig.~\ref{fig:4} plots the predicted and real differences between the WTP fetched by the producer's and the AI's games. 

Both amateur and professional producers initially predict that the AI-designed games are at a significant disadvantage, of \beliefDeficitAmCents{} and \beliefDeficitProCents{} cents respectively (\beliefDeficitAmSD--\beliefDeficitProSD{} SDs, one-sample $t$ test, $p\pBeliefDeficit$). The predicted gap turns in the AI's favor as soon as producers read the AI's idea, and stays there after they have played its game. This suggests that the AI outperforms humans in idea-generating. Further, this fact is plain for participants to see upon reading the AI's idea, even before experiencing the game or learning about market outcomes. However, it is not expected ex ante.

In answer to RQ2, the AI's advantage comes as a surprise to the market participants themselves. If such a bias in beliefs is prevalent outside our context, automation in creative tasks will meet producers who are unprepared for it. The relevant belief here is whether AI outperforms the producer themselves, not whether it outperforms a generic human producer. This belief may combine views about the technology with a tendency to overestimate one's own performance. Our experiment allows us to measure its bias because we observe each producer's actual and predicted performance relative to the AI.

\begin{figure}[t!]
\centering
\includegraphics[width=\textwidth]{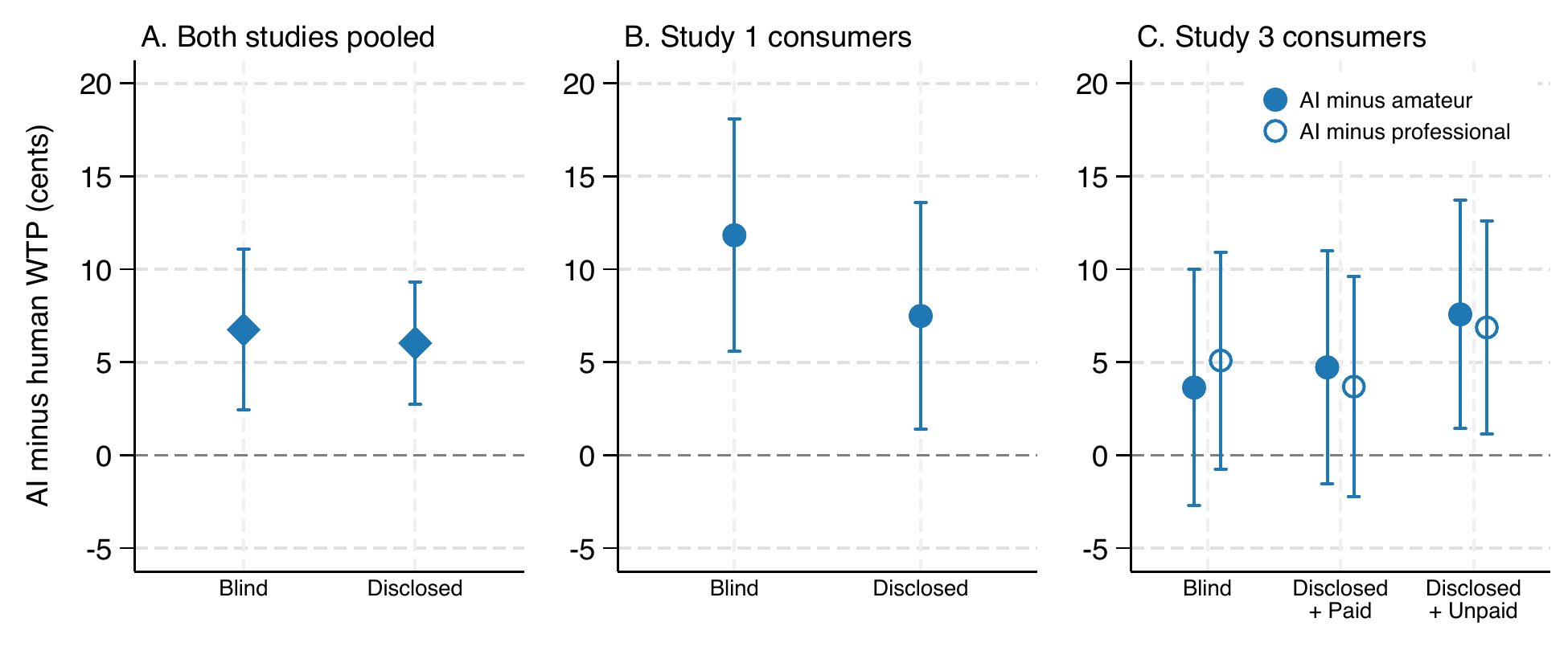}
\caption{\textbf{The AI willingness-to-pay premium by disclosure treatment.}
Difference in mean WTP between AI- and human-designed games by disclosure treatment: (\textbf{A}) pooling both studies, all human producers, and all Disclosed arms; (\textbf{B}) Study 1, where human games are amateur-designed; and (\textbf{C}) Study 3, separately for amateur (filled circles) and professional (hollow circles) games. Positive values indicate higher WTP for AI games. Error bars are 95\% confidence intervals, with standard errors two-way clustered by consumer and producer.}
\label{fig:5}
\end{figure}

\subhead{Consumers do not intrinsically prefer human-made products}
RQ3 asks whether, even if human-designed products are objectively of lower quality than those of the AI, consumers intrinsically prefer that humans have a hand in production.  Given the results above, this preference would have to be large enough to overcome the AI's advantage in taste for it to sustain human involvement in production. Our Blind and Disclosed arms test this hypothesis, and Fig.~\ref{fig:5} reports the advantage of AI-designed games in each of these treatments. Pooling across studies, types of producers, and the two Disclosed arms of Study 3 to maximize statistical power, we find no effect of disclosing a game's designer on consumers' WTP for it (Fig.~\ref{fig:5}A): the AI premium in the Blind arm is \pooledPremBlind{} cents ($p\pPooledPremBlind$) while in the Disclosed arm it is \pooledPremDisc{} cents ($p\pPooledPremDisc$), and we cannot reject that the AI premium is equal across arms ($p\pPooledBlindEqDisc$). That is, consumers prefer AI games even when the fact that no human was involved in producing them is made very salient. This result is reminiscent of evidence on human reactions to AI recommendations that demonstrates algorithm appreciation in actual, rather than hypothetical, choice \citep{logg2019algorithm}.

Disaggregating by study, we find no significant effect ($p\pBlindEqDisc$; table~\ref{tab:S4}) of disclosing a game's provenance on the difference between consumers' WTP for AI- and human-designed games in Study 1 (Fig.~\ref{fig:5}B). These results obtain despite several features of Study 1 that we might think push in favor of finding an intrinsic preference toward human-made products. Our consumers were themselves producers, having just designed games of their own to be sold in the same market. Consumers therefore knew the level of effort and input that production involved, and this kind of visibility ordinarily raises valuations of human-made products \citep{fuchs2015handmade, granulo2021symbolic, bellaiche2023humans}. Furthermore, we might have expected that the fact that consumers and producers were drawn from the same population would invite solidarity and other-regarding motives.

Though less precisely estimated, we also find no evidence that our Disclosed treatments in Study 3 affected consumers' relative WTP for human- vs.\ AI-designed games (joint $F$ test, $p\pArmsEqual$ for equality of the AI premium across the three arms; Fig.~\ref{fig:5}C). Not surprisingly then, there is also no difference between the disclosure treatment that announces that the human producer is paid the consumer's WTP and the treatment that announces that they are not. In addition, because Study 3 included consumers who never found themselves in the role of producers, we can rule out that the absence of an intrinsic preference for human-designed games in Study 1 merely reflected a sense of rivalry between producers.

A natural concern is that the null effects of disclosing provenance obtain only because consumers in our Blind arm are already able to tell which games were AI- and which were human-designed. To test this, in Study 3 we asked participants in the Blind treatment the probability that each game they faced was designed by a human. We find that consumers are wholly unable to distinguish the two, assigning a \sliderAI\% chance to AI games and a statistically indistinguishable \sliderHuman\% chance to human games ($p\pSlider$). This result is consistent with prior work showing that people cannot easily distinguish human- vs. AI-generated text \citep{jakesch2023human}. Finally, in both Disclosed treatments, we ask participants after providing their WTP whether they can recall whether the game was AI- or human-designed. Reassuringly, they answer this question correctly for \recallAcc\% of games, confirming that they attend to the information about who designed each game.

In answer to RQ3, we find no evidence that consumers intrinsically prefer human-made products. When we experimentally divulge the extent to which humans had a hand in production, their demand for the product remains unmoved and, in particular, they continue to significantly prefer AI to human products. 

\subhead{Exposure to AI production improves quality, but not pride or enjoyment}
We have seen that the AI outcompetes human producers, but we may also worry about secondary effects of AI exposure. On the one hand, people may be able to learn from the AI. On the other hand, they may take less pride in their work knowing it can be automated. We find suggestive evidence in favor of both channels. Recall that we randomize whether and when producers see the AI design a game. We therefore can estimate the effect of doing so on human production afterward, using OLS regressions of each outcome on an indicator for having watched the AI, with producer and task-order fixed effects (two-sided $t$ tests; Fig.~\ref{fig:S4}, table~\ref{tab:S5}). We find that human producers make better games after watching the AI produce; the games they produce become more fun ($p\pExpoFun$), more creative ($p\pExpoCreativity$), and more polished ($p\pExpoPolish$), though these translate into only an insignificant increase in their consumers' WTP (\expoWTP{} SDs, $p\pExpoWTP$; Fig.~\ref{fig:S4}A). But these changes do not lead to similar improvements in producers' attitudes toward their work. They predict no increase in WTP (\expoPred{} SDs, $p\pExpoPred$; Fig.~\ref{fig:S4}B), and if anything reduce their pride in their product: pride falls by \expoPride{} SDs ($p\pExpoPride$) and their WTP to include their game on a public website falls by \expoSite{} SDs ($p\pExpoSite$). We find no significant effects on enjoyment, WTP to be sent a link to their game, or producer surplus (all $p>0.10$).

\subsection*{Discussion}

We test two hypotheses for how a human role in economic production could persist despite growing AI capabilities and fail to find evidence for either: fully AI-produced games outperform those of amateurs and professionals alike, even when their (lack of) human involvement is revealed and made salient. Moreover, the AI's advantage is not anticipated by the human producers.

\subhead{Implications}
Our results suggest that AI could displace precisely the sort of work humans would most like to keep. The job in our experiment was a good one: producers took pride in their games and enjoyed designing them. Conventional measures would not register this loss. GDP records the consumption value of what is produced and the wages paid to produce it, but not the satisfaction and meaning workers derive from producing it. A world in which AI made our participants' games would therefore look like a richer one, even as something valued by the producers disappeared. Our producers expected their games to outsell the AI's and revised this belief only once they had seen the AI at work. Such surprise is likely to compound the personal costs of displacement: workers who do not anticipate being replaced have little reason to retrain or change occupations in advance, so their costs of adjusting are likely to be larger when replacement arrives. This speaks for frank and early public discussion of what AI production means for the labor market. Finally, our results offer little hope for strategies that emphasize a human hand in production. Labeling products as human-made did not raise what consumers were willing to pay for them, and neither did telling consumers that their payment would go to the human producer.

\subhead{Limitations}
No naturally occurring market lets us vary \emph{only} whether humans have a hand in making a product or what consumers know about it. A controlled experiment makes this variation possible, but poses the challenge of creating a paradigm where producers are familiar with and able to produce the product being sold and consumers actually demand the product. The production of computer games with a coding agent meets this challenge. Our market is nonetheless a compressed version of the markets it is meant to inform: prices were measured in small denominations, consumption lasted for a short period, and production lasted minutes rather than weeks. At the same time, many aspects of our design would seem to \textit{favor} human competitiveness: we handicapped AI-only production relative to what we allowed human designers, recruited a customer base unusually similar (indeed, identical) to and therefore likely sympathetic toward human producers, and made the level of AI involvement in production very salient. We expect our results to speak to creative work that sustains many mid-level jobs. It is less clear how far they extend to superstar markets, such as music or books, or to markets further afield, such as scientific research. Moreover, we cannot rule out that other interventions activate a preference for human production more effectively than disclosure, for example by creating familiarity between individual producers and consumers. Understanding the market forces likely to sustain or drive out humans from economic production in creative work is an urgent question for policy makers and social scientists. Our experiment provides early evidence and, we hope, motivates follow-up studies that address its limitations.

\clearpage
\renewcommand{\refname}{References and Notes}
\bibliographystyle{sciencemag}
\bibliography{bibliography}

\section*{Acknowledgments}

\paragraph*{Funding:} The authors received no external funding for this research.
\paragraph*{Author contributions:} J.J.C.\ and P.S.\ jointly designed the experiment, conducted the research, analyzed the data, and wrote the paper.
\paragraph*{Competing interests:} There are no competing interests to declare.

\subsection*{Supplementary materials}
Materials and Methods\\
Figs.\ S1 to S4\\
Tables S1 to S5\\
References \textit{(35--36)} %

\ifSM

\clearpage
\setcounter{figure}{0}\renewcommand{\thefigure}{S\arabic{figure}}
\setcounter{table}{0}\renewcommand{\thetable}{S\arabic{table}}
\setcounter{page}{1}\renewcommand{\thepage}{S\arabic{page}}

\begin{center}
{\Large\bfseries Supplementary Materials for}

\medskip

{\large Creative destruction: Human-AI competition in an experimental market}

\medskip

John J. Conlon and Peter Schwardmann\\
Correspondence to: \texttt{conlon.26@osu.edu}
\end{center}

\bigskip

\noindent\textbf{This PDF file includes:}\\
Materials and Methods\\
Figs.\ S1 to S4\\
Tables S1 to S5

\bigskip

\section*{Materials and Methods}

\subsection*{Participants and logistics}
Study 1's 296 US-based participants were recruited from Prolific between 6 and 10 July 2026. Participants who answered more than three comprehension questions incorrectly were excluded automatically by being asked to return the study early, which removed 9.7\% of potential participants. Table~\ref{tab:S1} reports summary statistics. As is common on Prolific, the sample skews somewhat more female, younger, better educated, and more liberal than the US population. Large majorities report using AI tools (\aiMonthly\%) and playing video games, including phone games, at least monthly (\gamesMonthly\%), and \aiCoding\% report using AI for coding or software outside the experiment. The median study duration was \medDurOne{} minutes. All three studies were preregistered on the AEA RCT Registry (AEARCTR-0019010) and approved by Carnegie Mellon University's Institutional Review Board (IRBSTUDY2016\_000000482); all participants gave informed consent. Screenshots of every page of all three instruments are available \href{https://johnjconlon17.github.io/website/Conlon_Schwardmann_2026_instrument_screenshots.pdf}{\textcolor{blue}{\underline{at this link}}}.

\subsection*{Elicitation of willingness to pay}
All WTP questions use a Becker-DeGroot-Marschak mechanism \citep{becker1964measuring}. A random price is drawn and displayed by a spinner landing on a value between \$0.00 and \$1.00; the participant pays the price and receives the good if the price falls below her stated WTP (framed as her ``maximum price''), and otherwise pays nothing and receives the default option. Payments are deducted from bonuses earned later in the study. 

\subsection*{Demo phase}
Participants are first introduced to the types of games involved in the experiment. There are seven ``basic games'': simplified black-and-white versions of well-known genres (falling blocks to be arranged into rows, a paddle that bounces a ball, and so on). Each participant is shown two randomly selected basic games, must play each for at least ten seconds, and then states a WTP to play one rather than the other for at least one minute. This phase familiarizes participants with the goods and with the WTP elicitation mechanism before they enter the production or consumption phases.

\subsection*{Producer phase}
Participants design three computer games with an LLM coding agent powered by Anthropic's Opus 4.8, accessed live during the experiment through the API. Each task begins from a basic game drawn at random from those not used in the demo phase. Participants send messages to the coding agent, which edits the game's code in response; before each new message they must test the current version by playing it for at least ten seconds. Design ends after three rounds of edits, and participants know this limit in advance. Figure~\ref{fig:2} shows the production interface: the live game appears on the right, the message history and elicitations on the left.

\textit{Predicting consumer WTP.} Producers know their finished game will be offered to another participant, who may pay up to \$1.00 to play it rather than the basic game the task began from, and that they earn that WTP as a bonus. They predict their consumer's WTP twice, once after seeing the basic game but before designing, and once after the final edit, earning \$0.10 for each prediction within \$0.02 of the realized WTP.

\textit{Producer surplus.} After testing the basic game but before writing their first message to the coding agent, producers state their WTP to design the game rather than skip the task, in which case the game is not offered to any consumer and the task pays no bonus. This is our measure of \textit{ex ante} producer surplus: the price a producer will pay to hold the job rather than skip it, which captures both the payment they expect from consumers and any non-pecuniary value of designing. The elicitation is implemented with 1\% probability, shown by a spinner landing on ``Counts'' or ``Doesn't Count.'' An analogous question is repeatedly asked before and after each response from the coding agent. Each of these questions also counts with a 1\% probability. If a question counts, a price is drawn as for the other WTP questions: if her WTP exceeds the price, she pays it and continues designing; otherwise she stops designing, and the game is not offered to a consumer. In either case, she is not asked any further such questions during that design task.

\textit{Pride and enjoyment.} After the third edit, participants state two incentivized WTPs: for a customized link to their game, so that ``you can play it later or share it with friends,'' and for inclusion of their game on a public website searchable by all participants. Purchased links and the website are delivered at the end of the study and by message on Prolific. They then answer unincentivized slider questions on how much they enjoyed designing the game and how proud they are of it, with text updating as the slider moves (e.g., ``I am [not at all/slightly/moderately/very/extremely] proud of the game I designed''). 

\subsection*{AI-only production}
Two-thirds of participants are randomly selected to watch the AI design a game without human input; half of these do so after their first design task and half after their second, generating random variation in both whether and when a participant sees autonomous AI production. Participants are truthfully told an LLM (Sonnet 4.6) is asked to ``redesign [the basic game] to be much more creative, engaging, and fun to play. Feel free to build on the basic design or completely redesign it.'' The full prompt additionally supplies technical context---for example that another model will implement the description as a computer game played for about a minute in a frame with external Start/Restart controls---followed by the basic game's HTML, and asks for a single short paragraph of plain prose. The reply is sent to the same coding agent participants use. Participants watch the LLM ``type'' its idea into an interface resembling their own, must test the resulting game for at least 30 seconds, and predict the consumer's WTP for it three times: before seeing the idea, after seeing the idea but before implementation, and after playing the implemented game. Fig.~\ref{fig:S1} shows this phase, with an example idea and the game produced from it.

\subsection*{Consumer phase}
Each participant is then offered up to four games drawn from a single other participant's producer phase, matched in real time: finished games enter a pool, and consumers draw from the producer who most recently finished. Consequently \nGamesValuedOne{} of \nEligibleGamesOne{} games were shown to exactly one consumer; the remaining \nUnvaluedGamesOne{} had no consumer because their producers were among the last to take the study, and those producers were paid \$1.20 per unmatched game. Because producers generate slightly fewer than four games in expectation, some consumers see fewer than four: \nRatedfour{} participants valued four games, \nRatedthree{} three, \nRatedtwo{} two, \nRatedone{} one, and \nRatedzero{} (the first participant) none. For each game, the consumer states a WTP to play it rather than its associated basic game, then plays the chosen game for at least a minute.

\textit{Disclosure treatment.} We randomize at the participant level whether game origin is disclosed. In the Disclosed arm, consumers are truthfully told, both before each game and in a caption shown while they play, whether it ``was designed by another participant who worked with an AI, just like you did'' or ``was designed by an AI on its own---without a participant working with it like you did.'' In the Blind arm every game is described only as ``produced during another participant's survey.''

\subsection*{Study 2: professional producers}
Study 2 ran the producer half of the Study 1 design on 180 participants recruited from Prolific between 4 and 11 August 2026 through its occupational screeners: current game developers or designers, content creators, graphic designers, and employees of the video-game industry, excluding Study 1 participants. This experiment was largely identical to Study 1 except that the consumer phase was dropped. Professionals were paid a Study 3 consumer's stated WTP for each of their games as a bonus, mirroring how Study 1's producers were paid from Study 1's consumer phase. Table~\ref{tab:S1} reports summary statistics.

\subsection*{Study 3: dedicated consumers}
Study 3 ran the consumer half of the design on 549 new Prolific participants (31 August to 2 September 2026; median duration \medDurThree{} minutes) who had taken part in neither earlier study. Each consumer valued four games drawn from the following pool: all 503 games designed by the professionals, 388 games sampled at random from Study 1's participants, and all 297 AI-designed games from the two producer runs (25\% of the pool). No consumer saw two human-designed games built from the same basic game. 

Consumers were randomized with equal probability to three arms that differed only in one sentence shown before each game. In the Blind arm, consumers were told that the pool mixes games designed by participants working with an AI and games designed by an AI on its own. In the two Disclosed arms every game was labeled with its origin. In the Disclosed + Paid arm, consumers were truthfully told for human-designed games that their producer would receive their stated WTP as a bonus; in the Disclosed + Unpaid arm they were told (also truthfully) that their WTP would not affect what the producer earned because another consumer's already had. Blind-arm consumers were asked after rating each game how likely it was to have been designed by a human, on a 0--100 slider paid by a matching-probability mechanism \citep{karni2009mechanism}; Disclosed-arm consumers were instead asked, right after stating their WTP, to recall whether the game had been labeled human- or AI-designed.

\subsection*{Statistical analysis}
Unless otherwise noted, $p$ values are from two-sided $t$ tests comparing group means, with standard errors two-way clustered by consumer and producer to account for repeated observations of each. Each AI-designed game is its own producer cluster; in Study 1 the two dimensions coincide, since each consumer valued one producer's games and each producer's games went to one consumer. 
Quantile comparisons in Fig.~\ref{fig:S3} are differences between empirical quantiles within each study, with 95\% percentile intervals from 10,000 bootstrap samples that resample consumers. In Fig.~\ref{fig:4}, we use robust standard errors for predicted gaps and for amateurs' actual gaps. Summary indices of the four quality ratings standardize each rating and combine them with weights proportional to the row sums of their inverse covariance matrix \citep{anderson2008multiple}, then standardize the result. All statistical tests are two-sided.

\clearpage

\begin{figure}[H]
    \centering
    \includegraphics[width=0.85\textwidth]{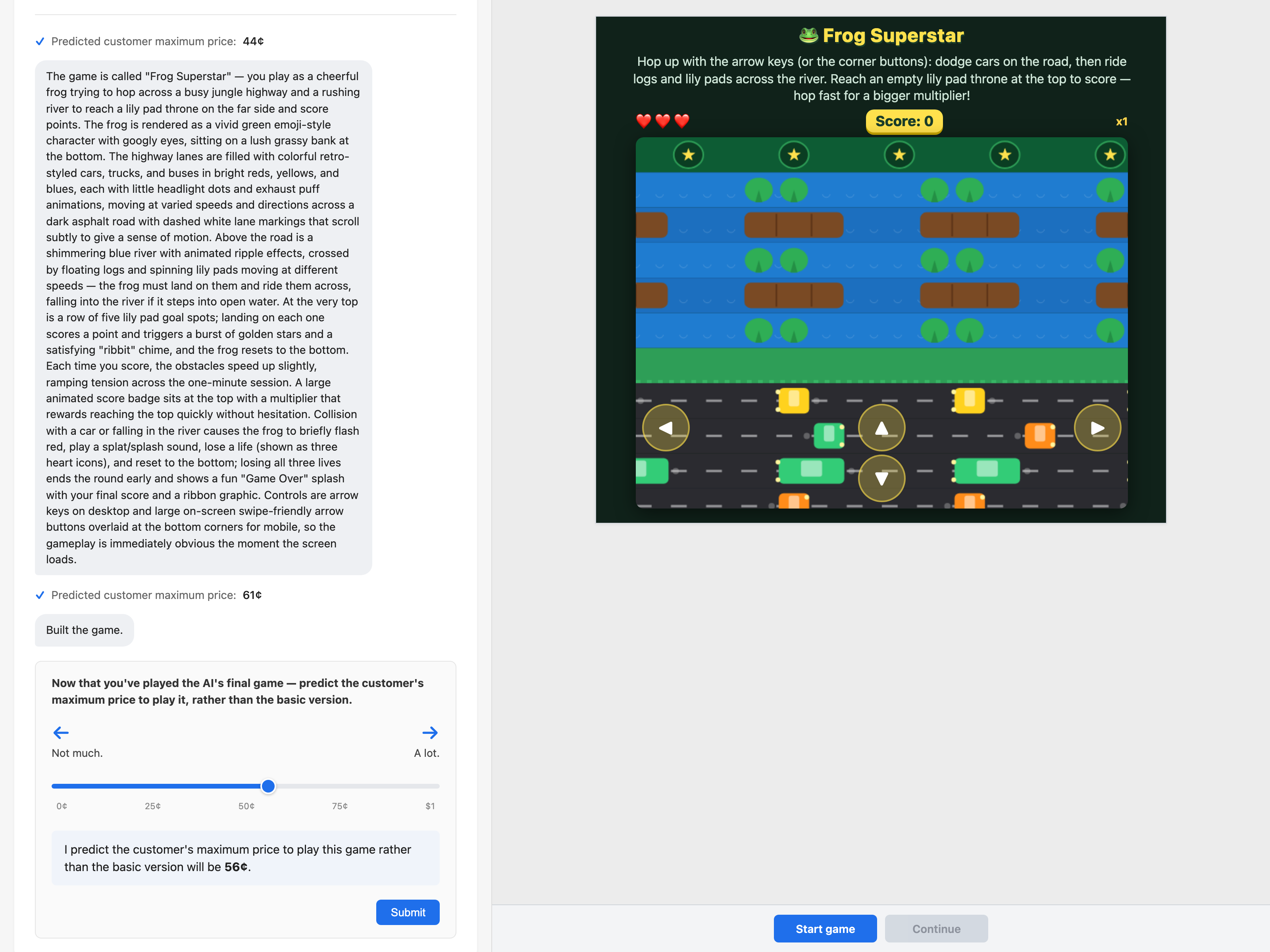}
    \caption{\textbf{AI-only production.} How autonomous AI production appears to participants, with an example of an LLM-generated idea and the game the coding agent produced from it.}
    \label{fig:S1}
\end{figure}

\begin{figure}[H]
\centering
\includegraphics[width=\textwidth]{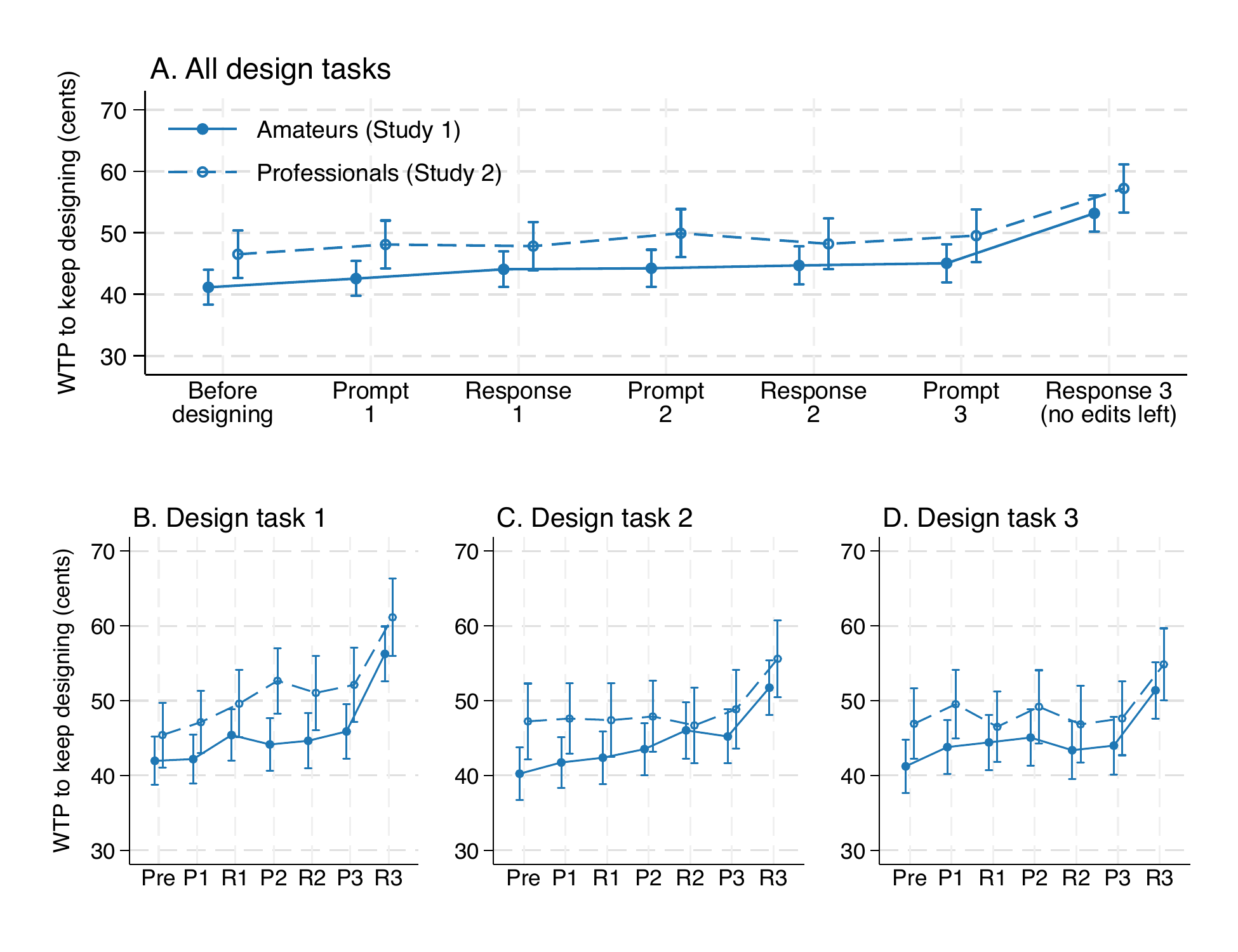}
\caption{\textbf{WTP to keep designing over the course of a design task.} Mean WTP to keep designing rather than give up designing at each of the seven elicitations within a design task: before designing (Pre), after writing each prompt (P1--P3), and after seeing each AI response (R1--R3). After the third response no edits remain, so the last elicitation is the WTP to offer the finished game to a consumer vs. not. (\textbf{A}) All design tasks; (\textbf{B} to \textbf{D}) design tasks 1 to 3. Filled circles denote amateur producers (Study 1); hollow circles denote professional producers (Study 2). The sample is the \nContTasks{} design tasks with all seven elicitations. Error bars are 95\% confidence intervals, with standard errors clustered by producer.}
\label{fig:S2}
\end{figure}

\begin{figure}[H]
\centering
\includegraphics[width=\textwidth]{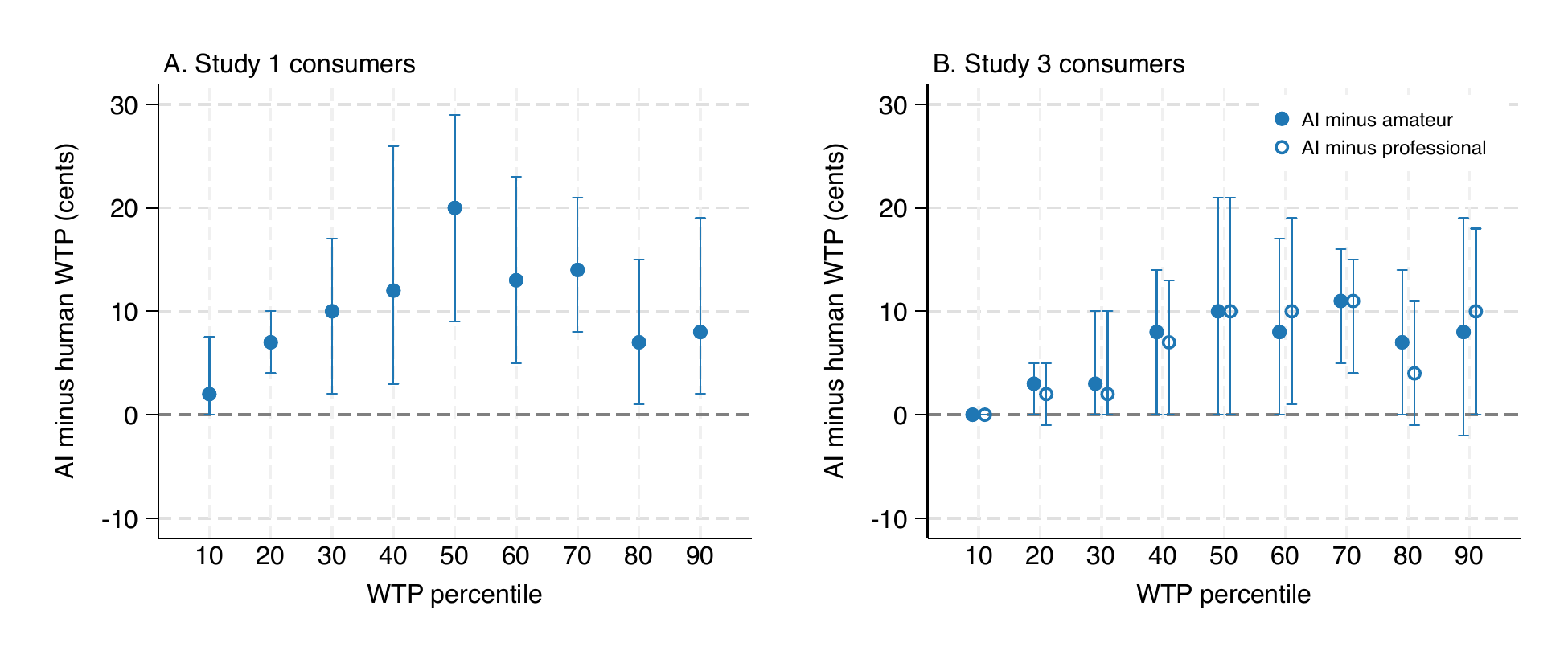}
\caption{\textbf{AI-minus-human willingness-to-pay gaps across quantiles.} AI-minus-human WTP gaps at the 10th through 90th percentiles in (\textbf{A}) Study 1 and (\textbf{B}) Study 3. Study 3 distinguishes AI-minus-amateur and AI-minus-professional gaps. Each point is the difference between AI- and human-designed games in WTP at a given percentile (for example, the 90th percentile of WTP for AI games minus the 90th percentile for human games). Error bars are 95\% percentile intervals from 10,000 bootstrap samples resampling consumers within each study.}
\label{fig:S3}
\end{figure}

\begin{figure}[H]
    \centering
    \includegraphics[width=\textwidth]{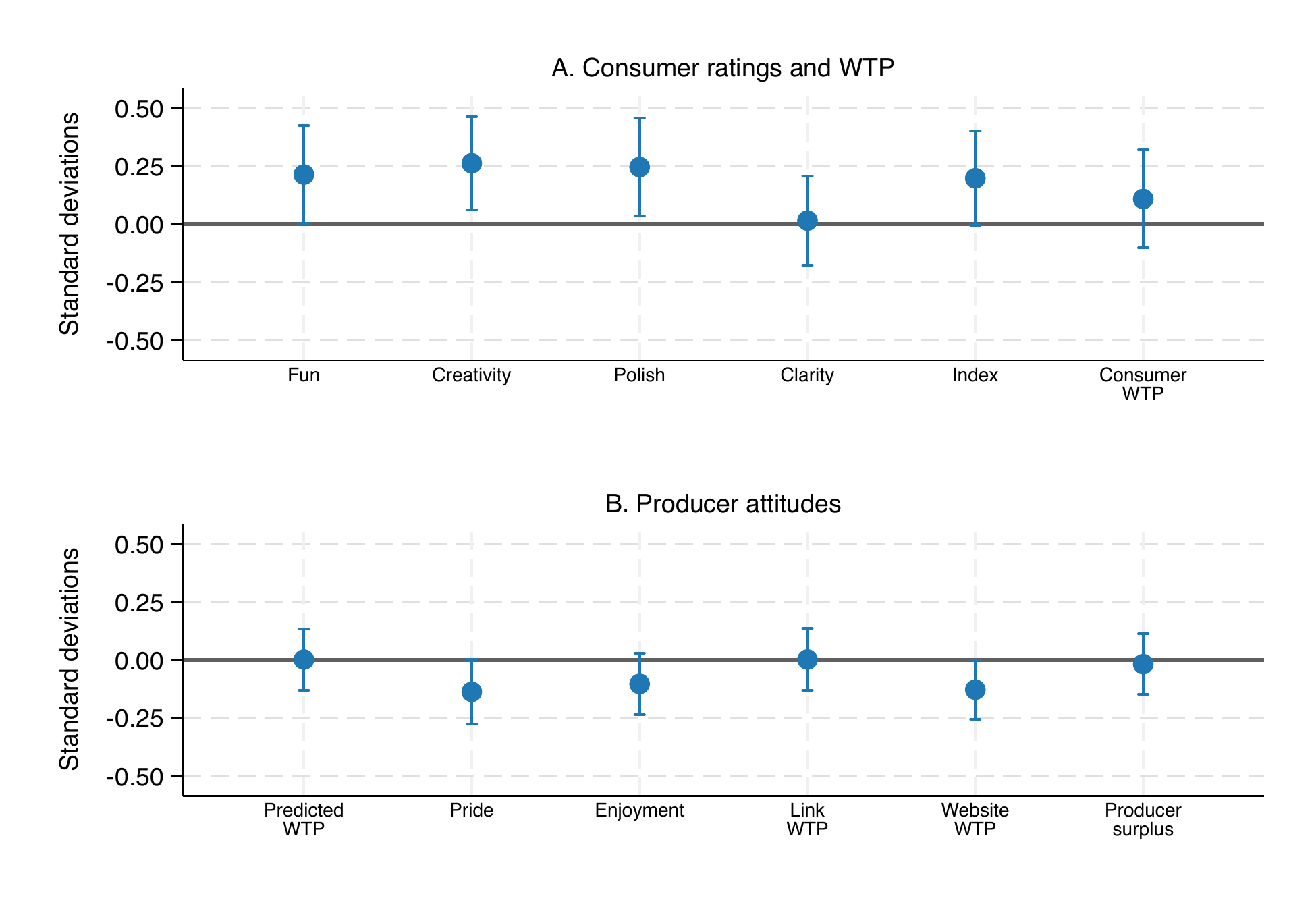}
    \caption{\textbf{Effect of having seen AI-only production on producer outcomes.} Each dot is the coefficient on an indicator for tasks completed after the producer watched the AI design a game on its own, from OLS regressions pooling Studies 1 and 2 with producer and task-order $\times$ study fixed effects. Outcomes are standardized to their own standard deviation, so coefficients are in standard-deviation units; error bars are 95\% confidence intervals. In (\textbf{A}), Fun, Creativity, Polish, and Clarity are Study 3 consumers' ratings of human-designed games, Index is their inverse-covariance-weighted summary~\citep{anderson2008multiple}, and Consumer WTP is those consumers' WTP for the game; each regression has one observation per consumer and game, with standard errors clustered by consumer and producer. In (\textbf{B}), Predicted WTP is the producer's final prediction of consumer WTP; Pride and Enjoyment are the unincentivized slider measures; Link WTP and Website WTP are the incentivized pride measures; Producer surplus is the initial WTP to design the game rather than skip it. Standard errors clustered by producer. Table~\ref{tab:S5} reports the underlying estimates.}
    \label{fig:S4}
\end{figure}

\clearpage

\begin{table}[H]\centering\small\linespread{1}\selectfont
  \begin{threeparttable}
    \caption{\textbf{Sample summary statistics.}}
    \label{tab:S1}
    \input{figures/tableS1.tex}
    \begin{tablenotes}
      \footnotesize
      \item \textit{Notes:} Study 1: the 296 amateur producers (Prolific, 6--10 July 2026). Study 2: the 180 professional producers (Prolific, 4--11 August 2026, recruited through occupational screeners). Study 3: the 549 consumers (Prolific, 31 August--2 September 2026).
    \end{tablenotes}
  \end{threeparttable}
\end{table}

\begin{table}[H]\centering\small\linespread{1}\selectfont
  \begin{threeparttable}
    \caption{\textbf{Consumer WTP and quality ratings (Study 3).}}
    \label{tab:S2}
    \input{figures/tableS2.tex}
    \begin{tablenotes}
      \footnotesize
      \item \textit{Notes:} OLS regressions of consumer willingness to pay on their standardized ratings of each game's quality. Columns 1--4 include one dimension at a time; column 5 uses their inverse-covariance-weighted summary index~\citep{anderson2008multiple}. Consumer fixed effects included in all columns. Standard errors two-way clustered by consumer and producer. $^{*}$ $p<0.10$, $^{**}$ $p<0.05$, $^{***}$ $p<0.01$.
    \end{tablenotes}
  \end{threeparttable}
\end{table}

\begin{table}[H]\centering\small\linespread{1}\selectfont
  \begin{threeparttable}
    \caption{\textbf{Producer outcomes, Studies 1 and 2.}}
    \label{tab:S3}
    \input{figures/tableS3.tex}
    \begin{tablenotes}
      \footnotesize
      \item \textit{Notes:} Producer-side outcomes across all design tasks of Study 1's amateurs and Study 2's professionals. Panel A reports market outcomes, Panel B the non-pecuniary amenities of the producer job, and Panel C the incentivized measure of \textit{ex ante} producer surplus (the WTP to undertake the design task rather than skip it).
    \end{tablenotes}
  \end{threeparttable}
\end{table}

\begin{table}[H]\centering\small\linespread{1}\selectfont
  \begin{threeparttable}
    \caption{\textbf{Consumer willingness to pay, by disclosure arm (Study 1).}}
    \label{tab:S4}
    \input{figures/tableS4.tex}
    \begin{tablenotes}
      \footnotesize
      \item \textit{Notes:} OLS regressions of consumer willingness to pay on treatment dummies interacted with whether the game was designed by AI. Column 2 adds fixed effects for the consumer's round within the consumer phase (1st through 4th elicitation). Standard errors are two-way clustered by consumer and by producer, with each AI-designed game treated as its own producer cluster. $^{*}$ $p<0.10$, $^{**}$ $p<0.05$, $^{***}$ $p<0.01$.
    \end{tablenotes}
  \end{threeparttable}
\end{table}

\begin{table}[H]\centering\small\linespread{1}\selectfont
  \begin{threeparttable}
    \caption{\textbf{Effect of watching AI-only production on producer outcomes.}}
    \label{tab:S5}
    \input{figures/tableS5.tex}
    \begin{tablenotes}
      \footnotesize
      \item \textit{Notes:} OLS regressions pooling Studies 1 and 2, with producer and task-order $\times$ study fixed effects; each dependent variable is standardized to its own standard deviation. Panel A uses Study 3 consumers' ratings of and WTP for human-designed games, with standard errors clustered by consumer and producer; Index is the inverse-covariance-weighted summary of the four ratings~\citep{anderson2008multiple}. Panel B uses producer-task-level outcomes with standard errors clustered by producer: Predicted WTP is the producer's final prediction of consumer WTP; Link WTP and Website WTP are the incentivized pride measures; Producer surplus is the initial WTP to design the game rather than skip it. $^{*}$ $p<0.10$, $^{**}$ $p<0.05$, $^{***}$ $p<0.01$.
    \end{tablenotes}
  \end{threeparttable}
\end{table}

\fi

\end{document}

%% file: figures/numbers_paper.tex
\input{figures/numbers_tab_summary_stats.tex}
\input{figures/numbers_tab_producer_outcomes.tex}
\input{figures/numbers_fig_continuation_wtp.tex}
\input{figures/numbers_tab_wtp_ai_disclosure.tex}
\input{figures/numbers_fig_tab_showcase_effects.tex}
\input{figures/numbers_fig_wtp_quality.tex}
\input{figures/numbers_fig_beliefs.tex}
\input{figures/numbers_fig_disclosure.tex}
\input{figures/numbers_prose.tex}

%% file: figures/numbers_tab_summary_stats.tex
\newcommand{\medDurOne}{54}
\newcommand{\medDurThree}{20}
\newcommand{\aiMonthly}{83.4}
\newcommand{\gamesMonthly}{83.8}
\newcommand{\aiCoding}{19.3}
\newcommand{\nProGames}{51}
\newcommand{\nProGraphic}{70}
\newcommand{\nProContent}{59}
\newcommand{\creatorJobPct}{80}

%% file: figures/numbers_tab_producer_outcomes.tex
\newcommand{\shareWTPpos}{83.5}
\newcommand{\prideMod}{80.7}
\newcommand{\prideExt}{42.5}
\newcommand{\enjoyMod}{87.5}
\newcommand{\enjoyLoved}{48.6}
\newcommand{\linkPos}{71.5}
\newcommand{\sitePos}{81.0}
\newcommand{\surplusPos}{92.1}
\newcommand{\surplusMean}{0.41}

%% file: figures/numbers_fig_continuation_wtp.tex
\newcommand{\nContTasks}{1,294}

%% file: figures/numbers_tab_wtp_ai_disclosure.tex
\newcommand{\pBlindEqDisc}{=0.33}

%% file: figures/numbers_fig_tab_showcase_effects.tex
\newcommand{\expoWTP}{0.11}
\newcommand{\pExpoWTP}{=0.31}
\newcommand{\expoSite}{0.13}
\newcommand{\pExpoSite}{=0.05}
\newcommand{\expoPride}{0.14}
\newcommand{\pExpoPride}{=0.05}
\newcommand{\expoPred}{0.001}
\newcommand{\pExpoPred}{=0.99}
\newcommand{\pExpoFun}{=0.05}
\newcommand{\pExpoCreativity}{=0.01}
\newcommand{\pExpoPolish}{=0.02}
\newcommand{\postExpoGapOne}{0.21}
\newcommand{\pPostExpoGapOne}{<0.01}
\newcommand{\postExpoGapThree}{0.15}
\newcommand{\pPostExpoGapThree}{=0.01}

%% file: figures/numbers_fig_wtp_quality.tex
\newcommand{\WTPhuman}{0.37}
\newcommand{\WTPai}{0.47}
\newcommand{\AIpremCents}{10.0}
\newcommand{\AIpremSD}{0.31}
\newcommand{\pAIprem}{<0.001}
\newcommand{\aiProWTPsd}{0.16}
\newcommand{\pAiProWTP}{<0.01}
\newcommand{\aiProIndex}{0.21}
\newcommand{\pAiProIndex}{<0.001}
\newcommand{\aiAmIndex}{0.34}
\newcommand{\pAiAmIndex}{<0.001}
\newcommand{\pProAmIndex}{=0.02}

%% file: figures/numbers_fig_beliefs.tex
\newcommand{\beliefDeficitAmCents}{9}
\newcommand{\beliefDeficitAmSD}{0.3}
\newcommand{\beliefDeficitProCents}{17}
\newcommand{\beliefDeficitProSD}{0.5}
\newcommand{\pBeliefDeficit}{<0.001}

%% file: figures/numbers_fig_disclosure.tex
\newcommand{\pooledPremBlind}{6.8}
\newcommand{\pooledPremDisc}{6.0}
\newcommand{\pPooledPremBlind}{<0.01}
\newcommand{\pPooledPremDisc}{<0.001}
\newcommand{\pPooledBlindEqDisc}{=0.79}
\newcommand{\pArmsEqual}{=0.87}

%% file: figures/numbers_prose.tex
\newcommand{\nUniqueValuedGamesFloor}{1,600}
\newcommand{\nEligibleGamesOne}{1,023}
\newcommand{\nUnvaluedGamesOne}{5}
\newcommand{\nGamesValuedOne}{1,018}
\newcommand{\nRatedzero}{1}
\newcommand{\nRatedone}{3}
\newcommand{\nRatedtwo}{15}
\newcommand{\nRatedthree}{123}
\newcommand{\nRatedfour}{154}
\newcommand{\sliderAI}{59.3}
\newcommand{\sliderHuman}{58.2}
\newcommand{\pSlider}{=0.67}
\newcommand{\recallAcc}{92.8}
\newcommand{\wordsAm}{60}
\newcommand{\wordsPro}{89}
\newcommand{\wordsAI}{235}

%% file: figures/tableS1.tex
\begin{tabular}{@{}l cc cc cc@{}}
\toprule
 & \multicolumn{2}{c}{Study 1: amateurs} & \multicolumn{2}{c}{Study 2: professionals} & \multicolumn{2}{c}{Study 3: consumers} \\
\cmidrule(lr){2-3}\cmidrule(lr){4-5}\cmidrule(lr){6-7}
\multicolumn{7}{l}{\textbf{A. Sample}} \\
\addlinespace[2pt]
N & \multicolumn{2}{c}{296} & \multicolumn{2}{c}{180} & \multicolumn{2}{c}{549} \\
Median duration (min) & \multicolumn{2}{c}{54} & \multicolumn{2}{c}{50} & \multicolumn{2}{c}{20} \\
Median age (years) & \multicolumn{2}{c}{37} & \multicolumn{2}{c}{32} & \multicolumn{2}{c}{39} \\
Female & \multicolumn{2}{c}{52.7\%} & \multicolumn{2}{c}{48.9\%} & \multicolumn{2}{c}{50.6\%} \\
Male & \multicolumn{2}{c}{43.2\%} & \multicolumn{2}{c}{48.9\%} & \multicolumn{2}{c}{46.6\%} \\
Other / no answer & \multicolumn{2}{c}{4.1\%} & \multicolumn{2}{c}{2.2\%} & \multicolumn{2}{c}{2.7\%} \\
Any college degree ($\geq$ 4-year) & \multicolumn{2}{c}{64.2\%} & \multicolumn{2}{c}{65.0\%} & \multicolumn{2}{c}{55.9\%} \\
Postgraduate degree & \multicolumn{2}{c}{20.3\%} & \multicolumn{2}{c}{12.8\%} & \multicolumn{2}{c}{18.2\%} \\
Democrat & \multicolumn{2}{c}{41.2\%} & \multicolumn{2}{c}{42.2\%} & \multicolumn{2}{c}{43.4\%} \\
Republican & \multicolumn{2}{c}{34.1\%} & \multicolumn{2}{c}{7.2\%} & \multicolumn{2}{c}{29.9\%} \\
Uses AI for coding & \multicolumn{2}{c}{19.3\%} & \multicolumn{2}{c}{28.9\%} & \multicolumn{2}{c}{18.2\%} \\
\addlinespace
\multicolumn{7}{l}{\textbf{B. Frequency of AI and video-game use}} \\
\addlinespace[2pt]
 & AI & Games & AI & Games & AI & Games \\
\cmidrule(lr){2-2}\cmidrule(lr){3-3}\cmidrule(lr){4-4}\cmidrule(lr){5-5}\cmidrule(lr){6-6}\cmidrule(lr){7-7}
Monthly & 12.5\% & 18.6\% & 13.3\% & 16.1\% & 10.6\% & 18.6\% \\
Weekly & 34.8\% & 31.1\% & 20.0\% & 23.9\% & 28.6\% & 27.5\% \\
Daily & 36.1\% & 34.1\% & 56.1\% & 43.9\% & 45.4\% & 33.2\% \\
\bottomrule
\end{tabular}

%% file: figures/tableS2.tex
\def\sym#1{\ifmmode^{#1}\else\(^{#1}\)\fi}
\begin{tabular}{l ccccc}
\toprule
 & (1) & (2) & (3) & (4) & (5) \\
\midrule
Fun (SD) & 14.93\sym{***} &  &  &  & \\
  & (0.67) &  &  &  & \\
\addlinespace
Creativity (SD) &  & 11.78\sym{***} &  &  & \\
  &  & (0.70) &  &  & \\
\addlinespace
Polish (SD) &  &  & 12.78\sym{***} &  & \\
  &  &  & (0.67) &  & \\
\addlinespace
Clarity (SD) &  &  &  & 12.17\sym{***} & \\
  &  &  &  & (0.70) & \\
\addlinespace
Summary index (SD) & & & & & 15.40\sym{***} \\
 & & & & & (0.64) \\
\addlinespace
Consumer FE & \checkmark & \checkmark & \checkmark & \checkmark & \checkmark \\
Observations & 2,196 & 2,196 & 2,196 & 2,196 & 2,196 \\
\bottomrule
\end{tabular}

%% file: figures/tableS3.tex
\begin{tabular}{@{}l cc @{\hspace{2.5em}} l cc@{}}
\toprule
 & Study 1 & Study 2 & & Study 1 & Study 2 \\
 & (amateurs) & (professionals) & & (amateurs) & (professionals) \\
\midrule
\multicolumn{6}{@{}l}{\textbf{A. Market outcomes}} \\
\addlinespace[2pt]
\multicolumn{3}{@{}l}{Consumer WTP received} & \multicolumn{3}{l}{Predicted consumer WTP} \\
\hspace{1em}Mean & \$0.37 & \$0.35 & \hspace{1em}Mean & \$0.52 & \$0.55 \\
\hspace{1em}Positive & 83.5\% & 80.0\% & \hspace{1em}Positive & 97.0\% & 96.4\% \\
\addlinespace
\multicolumn{6}{@{}l}{\textbf{B. Non-pecuniary amenities}} \\
\addlinespace[2pt]
\multicolumn{3}{@{}l}{Pride} & \multicolumn{3}{l}{Enjoyment} \\
\hspace{1em}Not at all & 9.4\% & 12.0\% & \hspace{1em}Not at all & 4.2\% & 5.9\% \\
\hspace{1em}Slightly & 9.9\% & 10.0\% & \hspace{1em}Slightly & 8.2\% & 6.5\% \\
\hspace{1em}Moderately & 16.3\% & 20.4\% & \hspace{1em}Moderately & 14.5\% & 17.1\% \\
\hspace{1em}Very & 21.9\% & 24.1\% & \hspace{1em}A lot & 24.4\% & 20.4\% \\
\hspace{1em}Extremely & 42.5\% & 33.5\% & \hspace{1em}Loved & 48.6\% & 50.2\% \\
\addlinespace
\multicolumn{3}{@{}l}{Link WTP} & \multicolumn{3}{l}{Website WTP} \\
\hspace{1em}Mean & \$0.32 & \$0.32 & \hspace{1em}Mean & \$0.36 & \$0.38 \\
\hspace{1em}Positive & 71.5\% & 72.9\% & \hspace{1em}Positive & 81.0\% & 83.7\% \\
\addlinespace
\multicolumn{6}{@{}l}{\textbf{C. Producer surplus}} \\
\addlinespace[2pt]
\multicolumn{3}{@{}l}{WTP to design} & \multicolumn{3}{l}{Design tasks} \\
\hspace{1em}Mean & \$0.41 & \$0.46 & \hspace{1em}N & 888 & 540 \\
\hspace{1em}Positive & 92.1\% & 93.2\% & & & \\
\bottomrule
\end{tabular}

%% file: figures/tableS4.tex
\def\sym#1{\ifmmode^{#1}\else\(^{#1}\)\fi}
\begin{tabular}{l cc}
\toprule
 & (1) & (2) \\
\midrule
AI $\times$ Blind & 11.83\sym{***} & 11.80\sym{***} \\
  & (3.18) & (3.19) \\
\addlinespace
AI $\times$ Disclosed & 7.49\sym{**} & 7.08\sym{**} \\
  & (3.09) & (3.08) \\
\addlinespace
Disclosed & -3.39 & -3.36 \\
  & (2.96) & (2.97) \\
\midrule
$p$-value: AI $\times$ Blind $=$ AI $\times$ Disclosed & 0.328 & 0.287 \\
\midrule
Order FE & -- & \checkmark \\
\midrule
Observations & 1018 & 1018 \\
\bottomrule
\end{tabular}

%% file: figures/tableS5.tex
\def\sym#1{\ifmmode^{#1}\else\(^{#1}\)\fi}
\begin{tabular}{@{}l *{6}{>{\centering\arraybackslash}p{1.35cm}}@{}}
\multicolumn{7}{@{}l}{\textbf{Panel A: Consumer ratings and WTP}} \\[2pt]
\toprule
 & Fun & Creativity & Polish & Clarity & Index & Consumer WTP \\
\midrule
Post AI-only game & 0.214\sym{**} & 0.263\sym{**} & 0.246\sym{**} & 0.016 & 0.198\sym{*} & 0.109 \\
  & (0.108) & (0.103) & (0.108) & (0.098) & (0.103) & (0.108) \\
\midrule
Task-order $\times$ study FE & \checkmark & \checkmark & \checkmark & \checkmark & \checkmark & \checkmark \\
Producer FE & \checkmark & \checkmark & \checkmark & \checkmark & \checkmark & \checkmark \\
\midrule
Observations & 1,600 & 1,600 & 1,600 & 1,600 & 1,600 & 1,600 \\
\bottomrule
\end{tabular}
\par\vspace{8pt}
\begin{tabular}{@{}l *{6}{>{\centering\arraybackslash}p{1.35cm}}@{}}
\multicolumn{7}{@{}l}{\textbf{Panel B: Producer attitudes}} \\[2pt]
\toprule
 & Predicted WTP & Pride & Enjoyment & Link WTP & Website WTP & Producer surplus \\
\midrule
Post AI-only game & 0.001 & -0.138\sym{*} & -0.104 & 0.001 & -0.129\sym{**} & -0.019 \\
  & (0.067) & (0.071) & (0.068) & (0.068) & (0.065) & (0.066) \\
\midrule
Task-order $\times$ study FE & \checkmark & \checkmark & \checkmark & \checkmark & \checkmark & \checkmark \\
Producer FE & \checkmark & \checkmark & \checkmark & \checkmark & \checkmark & \checkmark \\
\midrule
Observations & 1,321 & 1,354 & 1,354 & 1,354 & 1,354 & 1,411 \\
\bottomrule
\end{tabular}